\documentclass{article}
\usepackage[american]{babel}
\usepackage[backend=biber,style=nature]{biblatex}
\usepackage{todonotes}
\usepackage{grffile}
\usepackage{booktabs}
\usepackage{hyperref}
\usepackage{graphicx}
\usepackage{fullpage}
\usepackage{amsmath}
\usepackage{csquotes}
\usepackage{pdfpages}
\usepackage{placeins}

\renewcommand{\vec}[1]{\mathbf{#1}}

\title{Automatic discovery of cell types and microcircuitry from neural connectomics}
\author{Eric Jonas$^1$, Konrad Kording$^{2, 3,4}$}

\begin{document}
\maketitle

\begin{small}
\begin{enumerate}
  \item Electrical Engineering and Computer Science, University of California, Berkeley
  \item Department of Physical Medicine and Rehabilitation,
    Northwestern University and Rehabilitation Institute of Chicago,
    Chicago, Illinois
  \item Department of Physiology, Northwestern University, Chicago, Illinois
  \item Department of Applied Mathematics, Northwestern University, Chicago, Illinois
\end{enumerate}
\end{small}

\begin{abstract}
  Neural connectomics has begun producing massive amounts of data,
  necessitating new analysis methods to discover the biological and
  computational structure. It has long been assumed that discovering
  neuron types and their relation to microcircuitry is crucial to
  understanding neural function. Here we developed a nonparametric
  Bayesian technique that identifies neuron types and microcircuitry
  patterns in connectomics data. It combines the information
  traditionally used by biologists, including connectivity, cell body
  location and the spatial distribution of synapses, in a
  principled and probabilistically-coherent manner. We show that the
  approach recovers known neuron types in the retina and enables
  predictions of connectivity, better than simpler algorithms. It also
  can reveal interesting structure in the nervous system of
  C. elegans, and automatically discovers the structure of a
  microprocessor.  Our approach extracts structural meaning from
  connectomics, enabling new approaches of automatically deriving
  anatomical insights from these emerging datasets.
\end{abstract}

\section*{Introduction}
Emerging connectomics techniques \autocite{Morgan2013,Zador2012}
promise to quantify the location and connectivity of each neuron
within a tissue volume. These massive datasets will far exceed the
capacity of neuroanatomists to manually trace small circuits, thus
necessitating computational, quantitative, and automatic methods for
understanding neural circuit structure.  The impact of this kind of
high-throughput transition has been seen before -- rise of sequencing
techniques necessitated the development of novel computational methods
to understand genomic structure, ushering in  bioinformatics
as an independent discipline \autocite{Koboldt2013}.

The brain consists of multiple kinds of neurons, each of which is
hypothesized to have a specific role in the overall
computation. Neuron types differ in many ways, e.g. chemical or
morphological, but they also differ in the way they connect to one
another. In fact, the idea of well defined, type-dependent local
connectivity patterns (microcircuits) has been prominent in many
areas, from sensory (e.g. retina, \autocite{Masland2001} to processing
(e.g. neocortex \autocite{Mountcastle1997}) to movement (e.g. spinal
cord) \autocite{Grillner2005}. These sorts of repeated computing
patterns are a common feature of computing systems, even arising in
human-made computing circuits. It remains an important challenge to
develop algorithms to use anatomical data, e.g. connectomics, to
automatically back out underlying microcircuitry.

The discovery of structure is a crucial aspect of network
science. Early approaches focused on global graph properties, such as
the types of scaling present in the network \autocite
{WattsStrogatz1998}.  While this approach leads to an understanding
of the global network, more recent work aims at identifying very small-scale
repeat patterns, or “motifs” in networks\autocite{Milo2002}. These motifs
are defined not between different node types, but rather represent repeated
patterns of topology. 

The discovery of structure in probabilistic graphs is a well-known
problem in machine learning. Commonly used algorithms include
community-based-detection methods \autocite{Girvan2002}, and
stochastic block models \autocite{Nowicki2001}.  While these
approaches can incorporate the probabilistic nature of neural
connections \autocite{Hill2012} they do not incorporate the additional
richer structure present in connectomics data -- the location of cell
bodies, the spatial distribution of synapses, and the distances
between neurons. Of particular importance is that the probability of
connections has a strong spatial component, a factor that is hard to
reconcile with many other methods. A model attempting to fully capture
the variation in the nervous system should take into account the broad
set of available features.

When it comes to neuroscience and other computing systems, we expect
patterns of connectivity much more complex than traditional motifs,
exhibiting a strong spatial dependence arising from the complex
genetic, chemical, and activity-based neural development processes.

To address these challenges, here we describe a Bayesian
nonparametric model that can discover circuit structure automatically
from connectomics data: the cell types, their spatial patterns of
interconnection, and the locations of somata and synapses. We show
that by incorporating this additional information, our model both
accurately predicts the connection as well as agrees
with human neuroanatomists as to the identification of cell types.  

We primarily focus on the recently-released mouse retina connectome
\autocite{Helmstaedter2013}, but additionally examine the C. elegans
connectome \autocite{White1986}, and then ``connectome'' of a
classical microprocessor \autocite{James2010}. Comparing the cell
types discovered by the algorithms with those obtained manually by
human anatomists reveals a high degree of agreement. We thus present a
scalable probabilistic approach to infer microcircuitry from
connectomics data available today and in the future. 

\begin{figure}
  \centering 
  \centerline{\includegraphics[width=183mm]{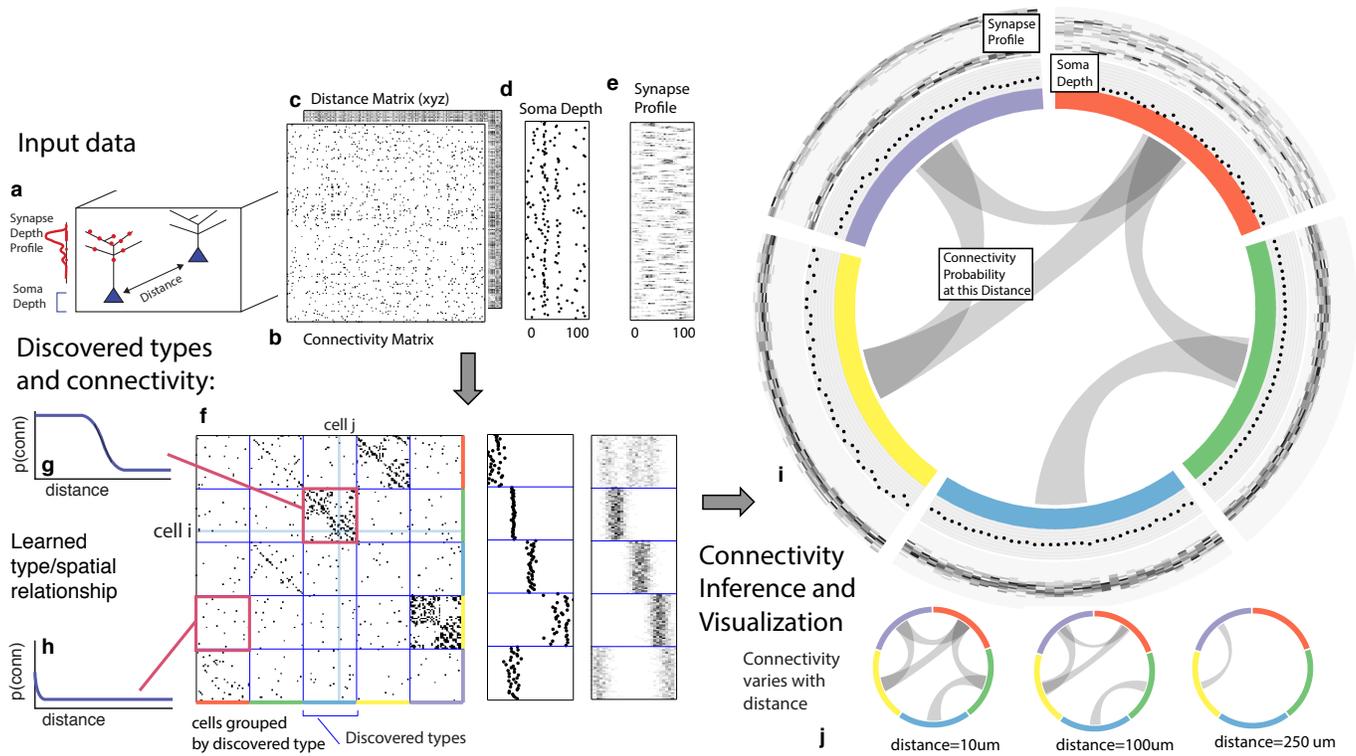}}
  \caption{Deriving circuitry and types from connectomics data a. As input we take the connectivity betwen cells (b), the distance between them (c), the depth of the cell bodies (d), and the depth profile of the syanpses (e). f. Our algorithm discovers hidden cell types in this
    connectivity data by assuming all cells of a type share a
    distance-dependent connectivity profile, similar depth, and a
    similar synpatic density profile, with cells of other types.  This
    results in a clustering of the cells by those hidden
    types. f.) shows the cell connectivity matrix with cells of the
    same type grouped together. g.) shows the learned probability of
    connection between our different types at various distances -- in
    this case, the cells are likely to connect when they are
    close. h. shows the probability of connection between two cell types
that very rarely connect -- there's a background ``base'' connection rate
to account for errors in data, but the probability is very low. 
We can plot this these types (i) to show the relationship and spatial
connections, and how probable various types are to connect to one
another. j.) The connectivity between our discovered types changes as a
function of distance. }

\label{fig:overview}
\end{figure}

\section*{Results}

We build a structured probabilistic model which begins with the
generic notion of a cell being a member of a single, unobserved type
-- and these types affect soma depth, distribution of synapses, as
well as a cell type and distance dependent connection probability. For
example, retinal ganglion cells may synapse on nearby, but not far
away, amacrine cells, with ganglion cells being superficial and having
a broad distribution of synapses.  

From these assumptions (priors) we develop a generative Bayesian model
that estimates the underlying cell types and how they connect. We take
as input (fig ~\ref{fig:overview}a) the connectivity matrix of cells (fig ~\ref{fig:overview}b) ,
a matrix of the distance between cells (fig ~\ref{fig:overview}c), the per-cell soma depth
(fig ~\ref{fig:overview}d) and the depth profile of the cell's
synapses (fig ~\ref{fig:overview}e). We perform joint probabilistic
inference to automatically learn the number of cell types, which cells
belong to which type, their type-specific connectivity, and how
connections between types vary with distance. We also simultaneously
learn the soma depth associated with each type and the typical
synaptic density profile (fig ~\ref{fig:overview}f-h.).

We apply our algorithm to datasets from mouse
retina, C. elegans and a historical microprocessor. Anatomists
classify cells based on many features, giving us a meaningful baseline
to compare against.

We start with a model for connectivity, the infinite stochastic block
model (iSBM)\autocite{Kemp2006a,Xu2006}, which has been shown to
meaningfully cluster connection graphs while learning the number of
hidden groups, or types. We extend this approach by adding distance
dependence to model salient aspects of microcircuitry via logistic and
exponential distance-link functions.  We additionally model cell body
depth unimodally and synapse density profile multimodially (see
Methods for mathematical details).

\begin{figure}
  \centering 
  \centerline{\includegraphics{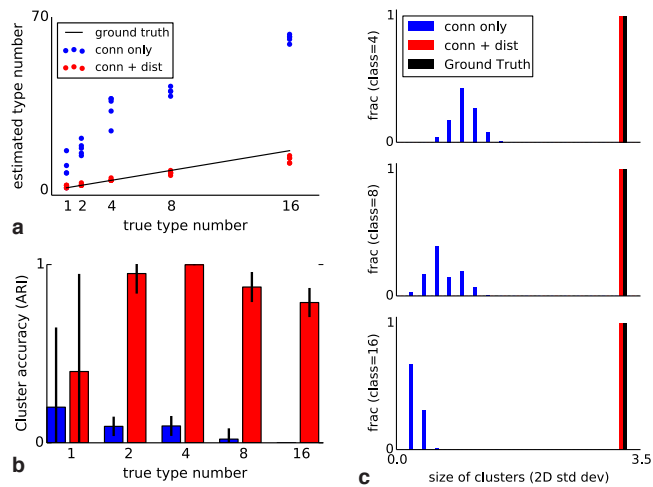}}
  \caption{Correct recovery of true numbers of hidden types in
    synthetic data when incorporating spatial information. a.) The
    infinite stochastic block model (which only uses connectivity
    information) over-estimates the number of classes as it fails to
    take distance into account, whereas our modeling of the
    combination of distance and connectivity finds close to the true
    number of classes. b.) ARI, a measure of ``correct clustering'',
    for different true class counts, between our model and a
    traditional iSBM. c.) For synthetic data, type discovery based
    only on connectivity are tiny spatially-localized groups -- in
    contrast, our model recovers the true spatial extent of the
    underlying types.  }
\label{fig:synthetic}
\end{figure}

To validate our model, we performed a series of simulations to test if
the model can accurately recover the true underlying network structure
and cell type identity.  We thus simulate data for which we know the
correct structure and comparing the estimated structure based on the
algorithm (see methods) with the one we used for simulation. We find
that the model does a good job of recovering the correct number of
cell types, (fig~\ref{fig:synthetic}a), the the cell identities
(fig~\ref{fig:synthetic}b), and the spatial extent of each type
(fig~\ref{fig:synthetic}c).  For comparison, existing infinite stochastic block model assumes cell type alone matters, and thus finds small neighborhoods of connected nodes (instead of global connectivity patterns). The model converges
relatively quickly to an estimate of the most probable values for the
cell types, which is enabled by using a combination of simulated
annealing and parallelized Markov chain Monte Carlo (see methods for
details). Thus we can apply our model to simulated datasets with structure
and scale similar to that of our biological datasets and recover the known 
correct structure. 

\subsubsection*{Learning types and circuitry in the retina}

\begin{figure}
  \centering 
  \centerline{\includegraphics{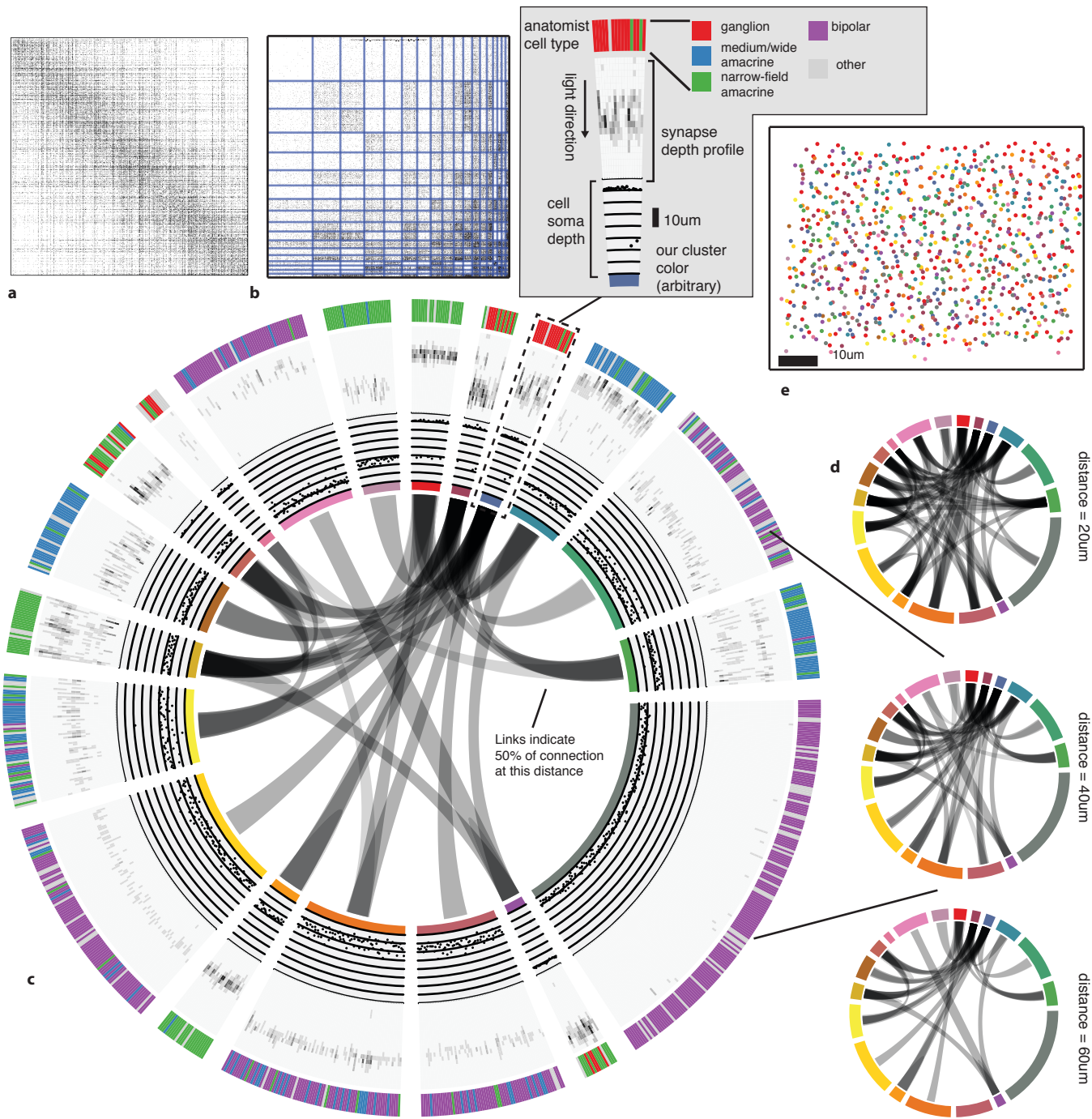}}
  \caption{Discovering cell classes in the mouse retina connectome.
    a.) Input connectivity data for 950 cells for which soma positions
    were known. b. clustered connectivity matrix. c.) connectivity
    diagram showing our clusters, as well as the cell depth and
    anatomist-labeled cell type of each cell. Our model shows moderate
    agreement with the cell types traditionally identified by
    anatomist laminar-specific connectivity patterns. 
    d.) Connectivity between our clusters as a function
    of distance -- the cluster consisting primarily of retinal
    ganglion cells (lower left-center) exhibits the expected near and
    far connectivity. e.)  The spatial distribution of our cell types
    -- each cell type tesselates space.}
\label{fig:mouseretina}
\end{figure}

The mouse retina \autocite{Masland2001} is a neural circuit which we
expect to have connectivity patterns that are well approximated by our
generative model. It is known that there are multiple classes of cells
that can be broadly grouped into: ganglion cells that transmit
information to the rest of the brain, bipolar cells that connect
between different cells, and amacrine cells that feed into the
ganglion cells. Recent research \autocite{Helmstaedter2013} has
produced a large dataset containing both the types of cells from
orthogonal approaches, and also the connectivity matrix between all
reconstructed cells.

The algorithm took less than 2 hours to perform inference, dividing
neurons into a set of cell types (fig~\ref{fig:mouseretina}c, each
wedge is a type). For each pair of neurons there is a specific
distance dependent connection probability
(fig~\ref{fig:mouseretina}b,c,d), which is well approximated by the
model fit. Moreover, each type of cell is rather isotropically
distributed across space (fig~\ref{fig:mouseretina}e) as should be
expected for true cell types.

Comparing the results of the algorithm to other information sources
allows evaluating the quality of the type determination. Our types
closely reflect the (anatomist-determined) segmentation of cells into
retinal ganglion, narrow amacrine, medium/wide amacrine, and bipolar
cells (fig~\ref{fig:mouseretina}c, outermost ring). We find that the
types we find tend to reflect the known laminar distribution in the
retina (fig~\ref{fig:mouseretina}c, middle ring). 

The algorithm yields a separation of neurons into a smaller number of
types than the fully granular listing of 71 types found by the
original authors of the paper, although is still highly correlated
with those finer type distinctions (see
supp~\ref{supp:mouseretina}. It is our expectation that, with larger
datasets, even closer agreement would be found.

\begin{figure}
  \centering 
  \centerline{\includegraphics{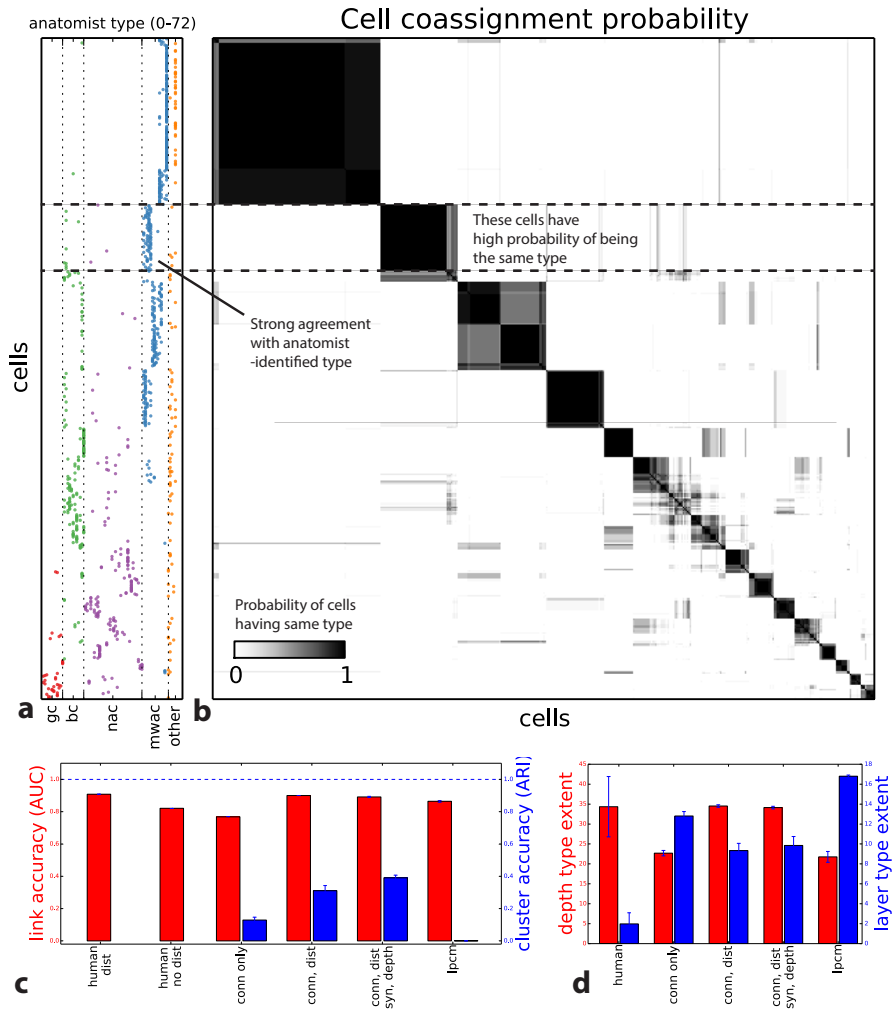}}
  \caption{\textbf{Visualizing type inference uncertainty.} Our fully Bayesian model gives a confidence estimate (posterior probability) that any two given cells are of the same type. In b.) we visualize that cell-cell coassignment matrix, showing the probability that cell i is of the same type as cell j on a range from 0.0 to 1.0. The block structure shows subsets of cells which are believed to all belong to the same type. For comparison, a.) shows the anatomist-defined type for each cell, grouped broadly into the coarse types identified in the previous figure. \textbf{Link vs cluster accuracy} c. A comparison of the
    predictive accuracy (area under the curve) for hand-labeled
    anatomical data, versus inclusion of additional sources of
    information, as well as the clustering accuracy. Note that our model sacrifies very little predictive accuracy for additional clustering accuracy. By comparison, conventional methods fail at one or both.  d. ) The spatial extent (in depth and area) of the
    types identified by humans and our various algorithmic
    approaches.}

\label{fig:mouseretina_compare}
\end{figure}

Our fully Bayesian model produces a distribution over probable
clusterings.  Figure~\ref{fig:mouseretina_compare} shows this
posterior distribution as a cell-cell coassignment matrix, sorted to
find maximum block structure. Each large, dark block represents a
collection of cells believed with strong probability to be of the same
type. When we plot (fig~\ref{fig:mouseretina_compare}b) the
anatomist-derived cell types along the left, we can see that each
block consists of a roughly-homogeneous collection of types. 

We evaluate our model along three sets of parameters
(Fig.\ref{fig:mouseretina_compare}): how closely does our clustering agree
with neuroanatomists' knowledge?  Given two cells, how accurately can
our model predict the link between them? And how closely does the
spatial extent (within a layer) of our identified types agree with the
known neuroanatomists.

As a first measure we compare link prediction accuracy across the
methods (Fig.\ref{fig:mouseretina_compare} B AUC, red).  We find that
given the dataset many techniques allow for good link-predictive
accuracy. All the methods allow decent link prediction with an AUC in
the .9 range. However, our algorithm clearly outperforms the simple
statistical models that only use connectivity.

As a second measure we compare link prediction accuracy across the
methods (Fig.\ref{fig:mouseretina_compare} B ARI, blue). We find that our
algorithm far outperforms the controls. We also find that when it is
based on more of the same information used by anatomists use then it
gets better at agreeing with these anatomists. In particular, using
connectivity, distance, synapse distribution and soma depth leads to
the highest ARI. When using the available information the algorithm
produces a good fit to human anatomist judgments.

Finally we look at the spatial extent of the discovered types both
within a layer and between layers (Fig.\ref{fig:mouseretina_compare}
C). We see that, in the absence of distance information, mere
connectivity information results in types which only span a small
region of space -- essentially “local cliques”. Incorporation of
distance information results in types which span the entire extent of
the layer. The depth variance of all models continues to be
substantially larger than that predicted by human anatomists -- future
directions of work include attempting to more strongly encode this
prior belief of laminarity.

\subsection*{Recovering spatial connectivity in multiple graphs simultaneously}

Having shown our model to work on the repeating tessellated, laminar structure of the mammilian retina, we then apply our model to a structurally very different connectome -- the whole body of a small roundworm: 
\textit{Caenorhabditis elegans} is a model system in
developmental neuroscience\autocite{White1986}, with the location and
connectivity of each of 302 neurons developmentally determined,
leading to early measurement of the connectome. Unlike the retina,
only the motor neurons in C. elegans exhibit regular distribution in
space -- along the body axis. Most interneurons are
concentrated in various ganglia that project throughout the entire
animal, and the sensory neurons are primarily located in a small
number of anterior ganglia. C. elegans also differs from the retina in that the measured connectome is actually two separate graphs -- one of directed chemical synapses and another of undirected electrical synapses. 

\begin{figure}
  \centering 
  \centerline{\includegraphics[width=4.5in]{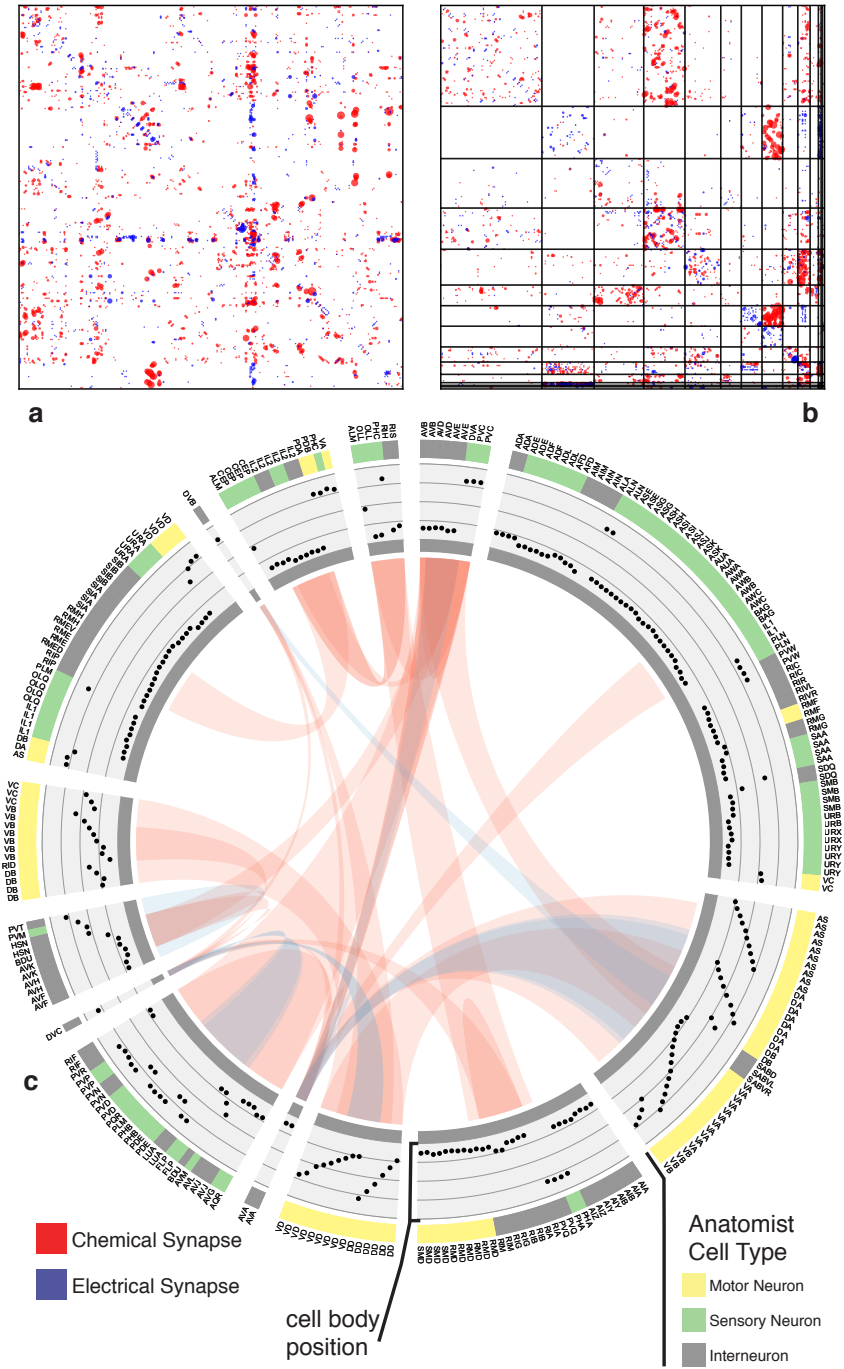}}
  \caption{Discovering connectivity and type in \textit{C. elegans}. 
 a.) Initial
    \textit{C. elegans} cell connectivity matrix, red are
    (directional) chemical synapses, blue are electrical
    synapses. Point size is proportional to synapse count between
    those cells. b.) Clustered adjacency matrix showing identified
    clusters. c.) Inter-cluster connectivity, showing the position of
    each cell along the body axis as well as the ``true'' cell type --
    both coarse (motor, sensory, and interneuron) as well as fine
    (small text labels). }
  \label{fig:celegans}
\end{figure}

Using both the chemical and electrical connectivity (see methods), we
determined the underlying clusters explained by connectivity and
distance (fig~\ref{fig:celegans}a). A superficial inspection of the results shows
clustering into groups consisting roughly homogeneously of motor
neurons, sensory neurons, and interneurons. Closer examination reveals
agreement with the classifications originally outlined by White in
1986.  

We identify cell types that reflect the known motor/non-motor neuron
classes, even though this system lacks the strong repeat
microcircuitry our model was designed for.  Motor-neuron types AS, DA,
and VA, all exclusively postsynaptic, are identified as a common type,
as are motor-neuron types VD and DD. Traditional types VC, DB, and VB
also mostly share a cluster. Various head motor neurons, including
types SMD and RMD, are clustered together. Interneurons with known
anatomically-distinct connectivity patterns, such as AVA (2 cells),
are clustered into pure types.The algorithm even correctly places the
single-cell types DVB and DVC by themselves.

Note our clustering does not perfectly reflect known divisions --
several combinations of head and sensory neurons are combined, and a
difficult-to-explain group of mostly VB and DB motor neuron types,
with VC split between various groups. Our identified cell types thus
reflect a ``coarsening'' of known types, based entirely on
connectivity and distance information, even when the organism exhibits
substantially less spatial regularity than the retina.

\subsection*{Types and connectivity in artificial structures}
\begin{figure}
  \centering 
  \centerline{\includegraphics[width=4in]{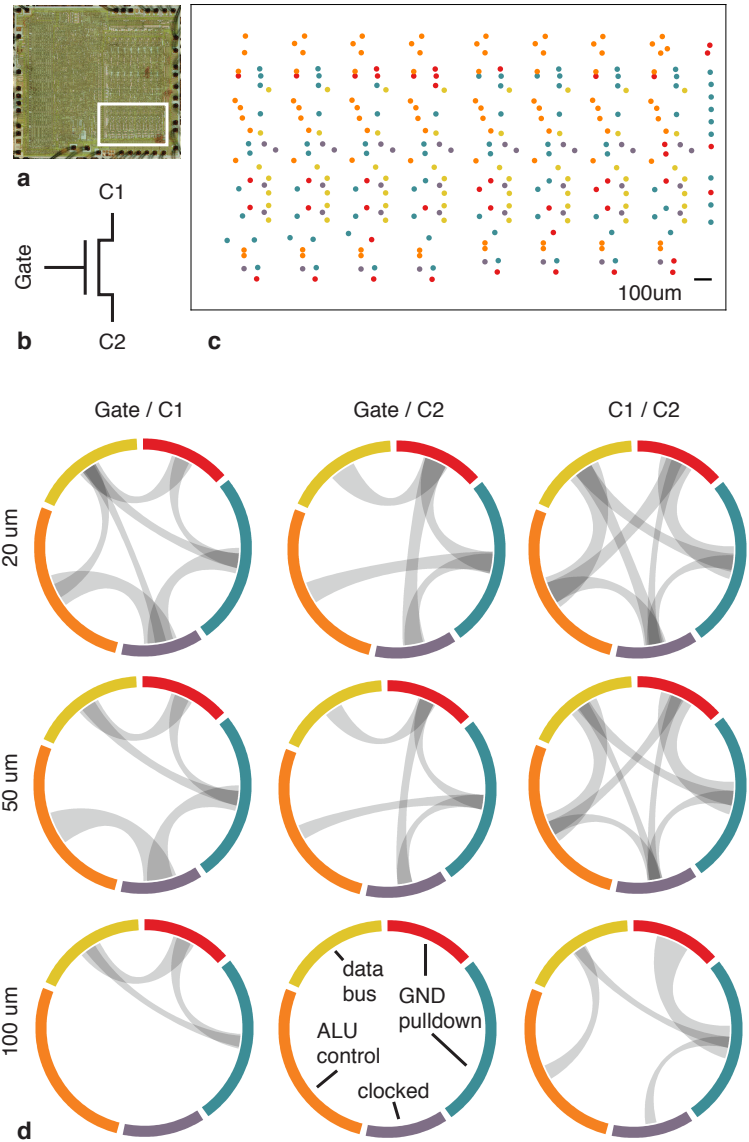}}
  \caption{Discovering connectivity and type the 6502 microprocessor.
    a.) is the micrograph of the original microprocessor, with the
    region containing the registers under study highlighted. b.) Our
    graph consists of the interconnections of MOS field-effect
    transistors with three terminals, Gate, C1, and C2. Reconstruction
    technique did not permit resolution of C1 and C2 into source and
    drain. c.) The spatial distribution of the transistors in each
    cluster show a clear pattern d.) The clusters and connectivity
    versus distance for connections between Gate and C1, Gate and C2,
    and C1 and C2 terminals on a transistor. Red and teal types have a
    terminal pulled down to ground and mostly function as
    inverters. Purple class are clocked, stateful transistors, orange
    control the ALU and yellow control the special data bus (SDB).}
  \label{fig:mos6502}
\end{figure}

To show the applicability of our method to other connectome-style
datasets, we obtained the spatial location and interconnectivity of
the transistors in a classic microprocessor, the MOS Technology 6502
(used in the Apple II) \autocite{James2010}. Computer architects use
common patterns of transistors when designing circuits, with each
transistor having a ``type'' in the circuit. We identified a region of
the processor with complex but known structure containing the primary
8-bit registers X, Y, and S (fig~\ref{fig:mos6502}).

Our algorithm identifies areas of spatial homogeneity that mirror the
known structure in the underlying architectural circuit, segmenting
transistor types recognizable to computer architects. Using the
original schematics, we see that one identified type contains the
``clocked'' transistors, which retain digital state. Two other types
contain transistors with pins C1 or C2 connected to ground, mostly
serving as inverters.  An additional identified type controls the
behavior of the three registers of interest (X, Y, and S) with respect
to the SB data bus, either allowing them to latch or drive data from
the bus. The repeat patterns of spatial connectivity are visible in
figure~\ref{fig:mos6502}c, showing the man-made horizontal and
vertical layout of the same types of transistors.

\section*{Discussion}
We have presented a machine learning technique that allows cell types
and microcircuitry to be discovered from connectomics data.  We have
shown its applicability to regularly structured laminar neural circuits like the retina, 
as well as a less structured whole neuronal organism (\textit{C. elegans}) and an artificial processor. When compared to existing methods, we show how the
incorporation of all of this data yields results that combine both
high link-prediction accuracy and high agreement with human
anatomists. We have found that combining the
available data types allows us to discover cell types and
microcircuitry that were known to exist in the systems based on
decades of previous research and allows good prediction of
connectivity.

For our probabilistic models, no known solution exists to
exactly find the most probable parsing of the neurons into cell-types
and connectivity patterns. We employ a collection of Markov-chain
Monte carlo techniques (see Methods) but while different
initializations converge to similar ultimate values, we can never
realistically obtain the global optimum. There are a broad range of
techniques that may offer good approximations to the global optimum
and future work could adapt them to
find more precise solutions to our problem.

For our probabilistic model, inference becomes slower as the amount of
data increases. Our algorithm required several hours for 1000
neurons. Scaling this class of probabilistic model is an active area
of research, and recent results in both variational methods
\autocite{Hoffman2013} and spectral learning \autocite{Anandkumar2012}
and future work could adapte them to find faster approximate solutions
to our problem.

Larger datasets will allow algorithms to distinguish more distinct
types and we expect closer agreement with existing anatomical
knowledge as more data become available.  Moreover, in general, for
such problems precision increases with the size of the dataset and the
cells that we have are not sufficient to statistically distinguish all
the cell types known in anatomy (such as the $\sim 70$ in the
retina). Still, using only connectivity and distance it is possible to
meaningfully divide neurons into types.

Our small collection of hand-selected distance-dependent likelihood
functions are clearly non-exhaustive, and assume monotonicity
of connectivity probability -- for a given class, closer cells
are never less-likely to connect. This is known to be insufficient
for various neural systems. Future models could incorporate
a wider variety of likelihood functions, or even learn the global
functional form from the data. 

There exist a range of previous approaches to the discovery of neural
microcircuitry\autocite{Mountcastle1957, Douglas1991, Bartho2004,
  Freund1998}.  These generally involve a great deal of manual labor
and ad-hoc determination of what constitutes a “type” of cell -- to
this day there are disagreements in the literature as to the “true”
types in the mammalian retina. Much as phylogenomics has changed our
understanding of animal ontologies, modern large scale data will allow
the efficient unbiased discovery of cell types and circuits. The sheer
amount of available data demands the introduction of algorithmic
approaches.

The development of automatic identification and quantification of cell
type may also provide a new computational phenotype for quantifying
the effect of disease, genetic interventions, and
developmentally-experienced neural activity. Our method can in
principle identify neuron-types across non-connected graphs,
e.g. across animals. For example, the types of neurons in one animal
can be associated with the types of neurons in another animal, in the
same way as this is already possible through molecular marker
\autocite{Brown2009}. This could be particularly important if cell
types appear that are due to properties of the stimuli and experience
as opposed to just the molecular properties of cells, such as color
and orientation selective types in primary visual cortex
\autocite{Sincich2005,Lennie2005}. This would allow comparative
quantitative anatomy across animals, and aid the search for the
ultimate causes of connectivity.

Our model combines connectivity, cellular, and synaptic properties,
and suggests the way towards combining even richer data. Distinct cell
types differ in morphology, connectivity, transcriptomics, relation to
behavior or stimuli and many other ways. Algorithms combining this
data and type type information may allow us to synthesize all the
available information from one experiment or even across experiments
into a joint model of brain structure and function.

Our work shows how rich probabilistic models can contribute to computational neuroanatomy. 
Eventually, algorithms will have to become a central tool for
anatomists, as it will progressively become impossible for humans to
parse the huge datasets. This transition may follow a similar
transition to that of molecular biology (with gene-finding
algorithms) and evolutionary biology with (computational
phylogenetics). Ultimately, computational approaches may help resolve the significant
disagreements across human anatomists.

\subsection*{Methods Summary}

For the basic link-distance model, we take as input a connectivity
matrix $R$ defining the connections between cell $e_i$ and $e_j$, as
well as a distance function $d(e_i, e_j)$ representing a (physical)
distance between adjacent cells.  See
the supplemental material for extension to multiple connectivity
matrices. We assume there exist an unknown number $K$ of latent
(unobserved) cell types, $k \in \{1, 2, 3, \dots, K\}$, and that each
cell $e_i$ belongs to a single cell type. We indicate a cell $e_i$ is
of type $k$ using the assignment vector $\vec(c)$, so $c_i = k$. The
observed connectivity between two cells $R(e_i, e_j)$ then depends
only on their latent type and their distance through a link function
$f(\cdot, d(e_i, e_j))$. We assume $f$ is parameterized based on the
latent type, $c_i=m$ and $c_j=n$, via a parameter$\eta_{mn}$, as well
as a set of global hyper parameters $\theta$, such that the link
function is $f(d(e_i, e_j) | \eta_{mn}, \theta)$.

We then jointly infer the maximum a posteriori (MAP) estimate of the
class assignment vector $\vec(c) = \{c_i\}$, the parameter matrix
$\eta_{mn}$, and the global model hyperparameters $\theta$ :

\begin{equation}
  p(\vec{c}, \eta, \theta | R ) \propto \prod_{i, j} p(R(e_i, e_j) | f(d(e_i, e_j) | \eta_{c_ic_j}), \theta) \prod_{m, n} p(\eta_{mn} | \theta)  p(\theta) p(\vec{c} | \alpha) p(\alpha) p(\theta)
\end{equation}

For the retina data, we then extend the model with the additional
features indicated. Cell soma depth is modeled as a
cell-type-dependent Gaussian with latent (unknown) per-type mean and
variance. Similarly, each cell has some number $N_i$ of synapses, 
each of which is drawn from a cell-type-specific density profile
with up to three modes.

Inference is performed in three steps via composable transition 
kernels -- one for structural, one for per-type parameters, and
one kernel for global parameters and hyperparameters. Details
of data preprocessing, inference parameters, and runtime can
be found in the Methods section. 

\printbibliography

\begin{itemize}

 \item \textbf{Acknowledgments} We thank Josh Vogelstein for discussions and reading of the manuscript, Finale Doshi-Velez for early discussions on the model, and Erica Peterson, and Jonathan Glidden, and Yarden Katz for extensive manuscript review. Funding for compute time was provided by Amazon Web Services ``AWS in Education'' grants. 
\item \textbf{Author Contributions} KK and EJ developed model. EJ derived inference, implemented code, tested, and ran experiments. KK and EJ wrote manuscript text and solicited feedback. 
 \item \textbf{Competing Interests} The authors declare that they have no
competing financial interests.
 \item \textbf{Correspondence} Correspondence and requests for materials
should be addressed to E.J.~(email: jonas@eecs.berkeley.edu).
\end{itemize}

\newpage
\section*{Methods }

\subsection*{Probabilistic Model}

Our model is a extension of the iSBM
\autocite{Kemp2006a,Xu2006} to incorporate spatial relations between entities,
inspired by attempts to extend these models with arbitrary
discriminative functions\autocite{Murphy2012}.

We take as input a connectivity matrix $R$ defining the connections
between cell $e_i$ and $e_j$, as well as a distance function $d(e_i,
e_j)$ representing a (physical) distance between adjacent cells. See
the supplemental material for extension to multiple connectivity
matrices. We assume there exist an unknown number $K$ of latent
(unobserved) cell types, $k \in \{1, 2, 3, \dots, K\}$, and that each
cell $e_i$ belongs to a single cell type. We indicate a cell $e_i$ is
of type $k$ using the assignment vector $\vec(c)$, so $c_i = k$. The
observed connectivity between two cells $R(e_i, e_j)$ then depends
only on their latent type and their distance through a link function
$f(\cdot, d(e_i, e_j))$. We assume $f$ is parameterized based on the
latent type, $c_i=m$ and $c_j=n$, via a parameter$\eta_{mn}$, as well
as a set of global hyper parameters $\theta$, such that the link
function is $f(d(e_i, e_j) | \eta_{mn}, \theta)$.

We then jointly infer the maximum a posteriori (MAP) estimate of the
class assignment vector $\vec(c) = \{c_i\}$, the parameter matrix $\eta_{mn}$, and
the global model hyperparameters $\theta$ :

\begin{equation}
  p(\vec{c}, \eta, \theta | R ) \propto \prod_{i, j} p(R(e_i, e_j) | f(d(e_i, e_j) | \eta_{c_ic_j}), \theta) \prod_{m, n} p(\eta_{mn} | \theta)  p(\theta) p(\vec{c} | \alpha) p(\alpha) p(\theta)
\end{equation}


We describe the spatial ``Logistic-distance Bernoulli''  function here,
and others in the supplemental material. 

The ``logistic-distance Bernoulli'' spatial model assumes that, if cell
$e_i$ is of type $m$ and cell $e_j$ is of type $n$, then $\eta_{mn}
= (\mu_{mn}, \lambda_{mn})$, and the probability that two cells $e_i$
and $e_j$ are connected is given by
\begin{eqnarray}
p^* &=& \frac{1.0}{1 + \exp \frac{d(e_i, e_j) - \mu_{mn}}{\lambda_{mn}}}\\
p &= & p^* \cdot (p_{max} - p_{min}) + p_{min}
\end{eqnarray}
where $p_{max}$ and $p_{min}$ are global per-graph parameters. 

We place an exponential priors on the latent parameters:
\begin{eqnarray}
 \mu_{mn} \sim \exp(\mu | \mu^{hp}) \\
\lambda_{mn} \sim \exp(\lambda | \lambda^{hp})
\end{eqnarray}

using  $\lambda^{hp}$ and $\mu^{hp}$ as global per-graph hyperparameters. 

We use a Dirichlet-process prior on class assignments, which allows
the number of classs to be determined automatically. In brief, for $N$
total cells, the probability of a cell belonging to a class is
proportional to the number of datapoints already in that class, $N_k$,
such that $p(c_i = k) \propto \frac{m_k}{N + \alpha}$ and the
probability of the cell belonging to a new class $k'$ is $p(c_i = k')
\propto \frac{\alpha}{N + \alpha}$. $\alpha$ is the global
concentration parameter -- larger values of $\alpha$ make the model
more likely to propose new classes. We grid the parameter $\alpha$ and
allow the best value to be learned from the data.

Where we model cell depth, we assume that 
each cell type has a typical depth, and thus a Gaussian distribution
of $s_i$. We assume $s_i \sim N(\mu^{(s)}_k,
\sigma^{2(s)}_k)$, where the $(s)$ superscript indicates
these model parameters are associated with the soma-depth 
portion of our model.  We use a conjugate prior for $(\mu^{(s)}_k, \sigma^{2(s)}_k)$ with
$\mu^{(s)}_k \sim N(\mu^{(s)}_{hp}, \sigma^{2(s)}_k/ \kappa^{(s)}_{hp})$ and $\sigma^{2(s)}_k \sim
\chi^{-1}(\sigma^{2(s)}_{hp}, \nu^{(s)}_{hp}$. The use of conjugacy simplifies inference while allowing for each cell-type to have its own depth mean and distribution. 

Where we model synapse depth profile, we assume that each cell
type has a characteristic depth distribution of synaptic contact points, and thus
a mixture of Gaussians distribution over cell $i$’s $N_i$ contact points, $\vec{g^i}$.
We do this by assuming the $g^i_j$
are drawn from an $M=3$-component mixture of Gaussians. Thus associated with each cell type $k$ is a vector
of $M$ Gaussian means $(\mu^g_{k,1}, \cdots, \mu^g_{k, M})$, and a
mixture vector $\pi_k$.  This representation can thus model depth distributions of contact points that have up to three modes, an assumption that is well matched in the bulk of anatomical studies of cell-type dependent connectivity.

\subsection*{Inference} 
We perform posterior inference via Markov-chain Monte Carlo (MCMC),
annealing on the global likelihood during the traditional burn-in
phase. MCMC transition kernels for different parts of the state space
can be chained together to construct a kernel whose ergodic
distribution is the target ergodic distribution over the entire state space. 

Our first transition kernel (``structural'') performs gibbs sampling 
of the assignment vector $p(\vec{c} | \eta, \theta, \alpha)$. 
The lack of conjugacy in our likelihood model makes an explicit 
evaluation of the conditional assignment probabilities impossible, 
motivating us to use an auxiliary variable method \autocite{Neal2000}
in which a collection of ephemeral classs are explicitly represented
for the duration of the Gibbs scan. 

We then employ a transition kernel to update the per-component
parameter values $\eta_{mn}$. Conditioned on the assignment vector
$\vec{c}$ and the model hyperparameters $\theta, \alpha$ the 
individual $\eta_{mn}$ are independent. We slice sample \autocite{Neal2003}
each component's parameters, choosing the slice width as a function
of the global hyperparameter range. 

The global hyper-parameters, both $\alpha$ and $\theta$, are allowed
to take on a discrete set of possible values. As $\theta$ is often a
tuple of possible values, we explore the cartesian product of all
possible values. We then Gibbs sample (our final transition kernel),
which is always possible in a small, finite, discrete state space.

We chain these three kernels together, and then globally anneal on the
likelihood from a temperature of $T=64$ down to $T=1$ over $900$
iterations unless otherwise indicated, and then run the chain for
another $100$ iterations. We then generate at least $20$ samples, each
taken from the end of a single Markov chain initialized from different
random initial points in the state space. For visualization we pick the
chain with the highest log likelihood, but for all numerical
comparisons (including link probability and cluster accuracy) we use this full
collection of samples from the posterior distribution to estimate the
resulting statistics.

\subsection*{Link Prediction}
As a proxy for link-prediction accuracy we compute the probability of a link
between two cells using each model, trained fully on the data. While this method
is potentially prone to overfitting, the overfitting will be shared across models
and in fact will preferentially bias in favor of competing models
which over-cluster the data. We use a full collection of posterior samples when
computing the link probability, and then compute the area under the ROC curve for
each. 

\subsection*{Model Comparison}
We compare our model with a standard network clustering model, the
latent-position clustering model. This model assumes each cell belongs
to one of K clusters, and each cluster is associated with a
$d-$dimensional Gaussian distribution. The probability of a link is
then a function of the distance between the data points in this
continuous-space. We use \autocite{Salter-Townshend2013} a variational
implementation provided in R, parametrically varying the number of
latent dimensions and the number of requested groups.  While this
model provides reasonable link predictive accuracy, the clusterings
dramatically disagree with those from human anatomists.

\subsection*{Parameters}

Hierarchical generative models can be sensitive to hyperparameter
settings, thus for most hyperparamters we perform inference. In cases
where we cannot we run separate collections of markov chains at
separate settings and show the results across all pooled
parameters. For the case of the mouse retina data, we consisder
maximum link probability $p_{max} \in \{0.95, 0.9, 0.7\}$, variance
scales for the synapse density profile of $\sigma^2 \in \{0.01, 0.1,
1.0\}$ (of normalized depth), and $K \in \{2, 3\}$ possible synapse
density profile mixture components. For the connectivity-distance-only
model we actually perform inference over both $p_{max}$ and $p_{min}$.

\subsubsection*{Mixing of our Markov chains} 
Evaluating whether or not approximate inference methods, such as MCMC,
produce samples which are valid approximations of the posterior
distribution is an ongoing area of research in the computational
statistics community. We use a rough proxy here -- synthetic
likelihood evaluation.  For synthetic datasets of sizes comparable to
our real data size, do we recover known ground truth information after
running our markov chains for the appropriate amount of time?

\begin{figure}
  \centering 
  \centerline{\includegraphics[width=60mm]{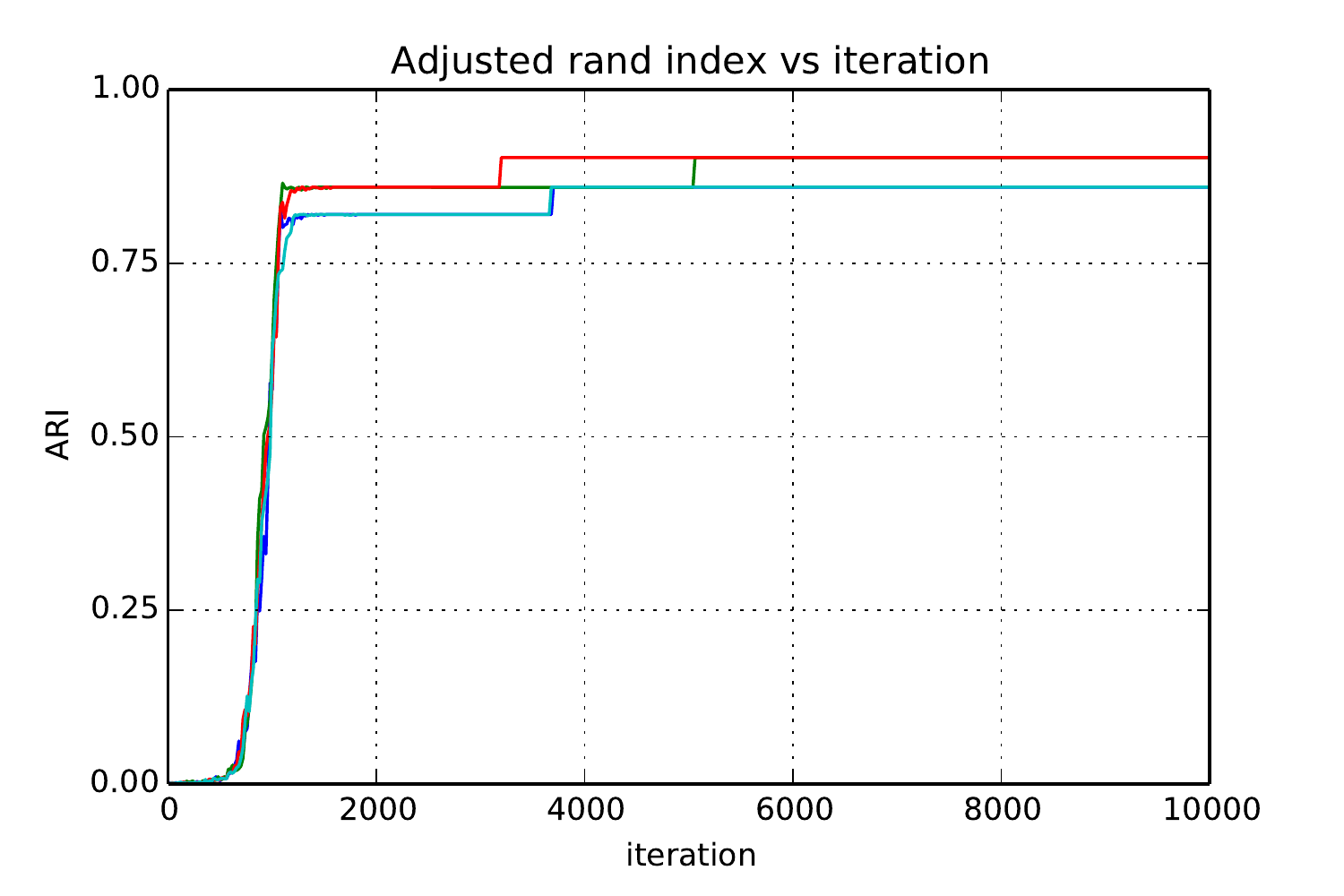}}
  \caption{Adjusted rand index for synthetic data as a function of run iteration. }
\label{fig:mixing:ari}
\end{figure}

\begin{figure}
  \centering 
  \centerline{\includegraphics[width=130mm]{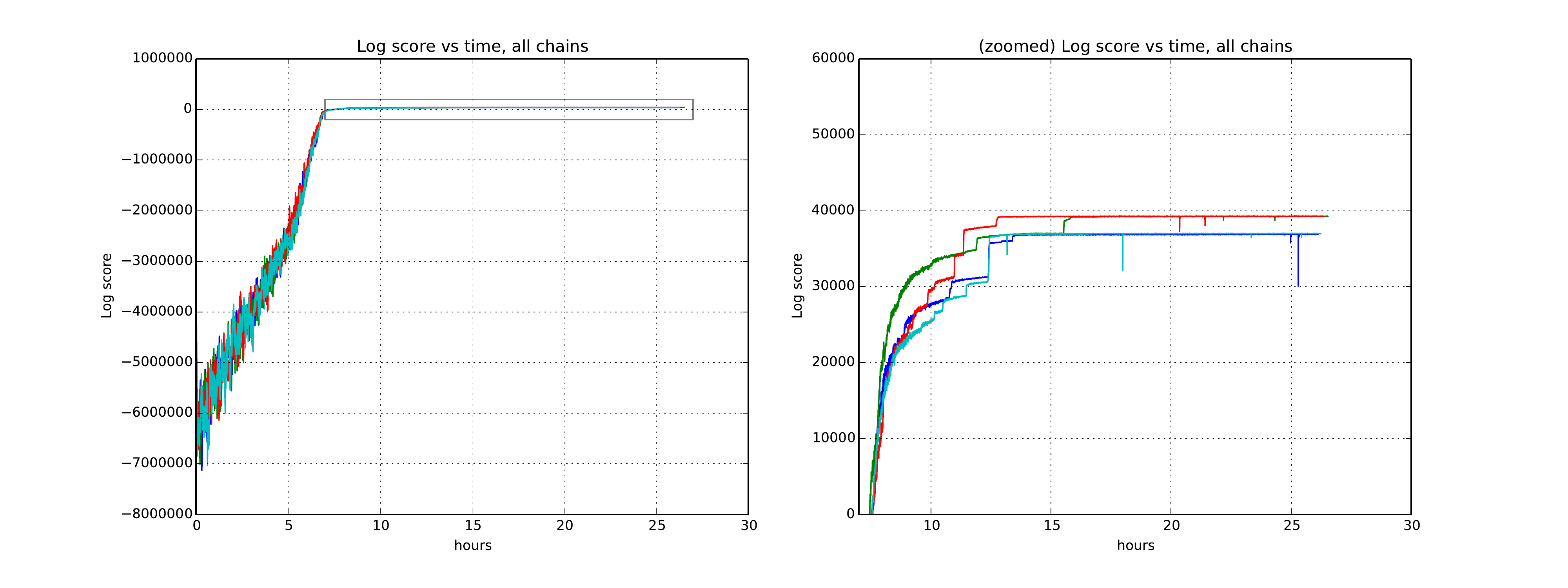}}
  \caption{Total model score (log score) vs iteration}
\label{fig:mixing:scorevstime}
\end{figure}

Figures \ref{fig:mixing:ari} and \ref{fig:mixing:scorevstime} shows
the cluster accuracy (ARI) to ground truth and the total log score as a function of
runtime.  We see dramatic changes in log score initially as we vary
the temperature, stabilizing as runtime progresses, for each
chain. Then we see the characteristic jumps between nearby modes
towards the end of the run, in both log score and ARI.  Importantly,
regardless of whether our model over- or under-estimates the exact
posterior variance about the network, we find points in the latent
variable space that are both predictive \textit{and} parsimonious,
largely agreeing with the human anatomists and predicting existing
connections.

\subsection* {Mouse Retina}
\label{supp:mouseretina}
Dense serial electron microscopy of a $114\mu m \times 80 \mu m $ area
in the mouse retina by \autocite{Helmstaedter2013} yielded a listing
of places where neurons come into contact. There were over 1000 cells
originally, and selected the $950$ for which the location of the soma
could be reconstructed from the provided cell plots (soma locations
were not provided by the study's authors in machine-readable
form). Ultimately this left a matrix between the total synapse-like
contact area between all pairs of 950 cells. Area was thresholded at
$0.1\mu m$, determined by hand, to yield a 950 $\times$ 950 entry
matrix that served as input to our algorithm. We measured the distance
between cells using the reconstructed soma centers, and used the
Logistic-Distance spatial relation. Hyperprior griddings are shown in
supplemental section \ref{supp:hyperpriors}.

\subsection*{C. elegans}

We obtained the connectome of c. elegans from data published
previously \autocite{Varshney2011}, and isolated the 279 nonpharyngeal
neurons, with a total of 6393 chemical synapses and 890 gap junctions
originally cleaned up in \autocite{Chen2006}. A cell's position was
its distance along the anterior-posterior axis normalized between zero
and one. We used both networks, the chemical network as a directed
graph and the electrical network as undirected graph. We use the
synapse counts with the logistic-distance poisson likelihood, scaling
the counts by 4.0 to compensate for the Poisson's overdispersion.

\subsection*{Microprocessor}
We extracted the connection graph for the transistors in the MOS6502
\autocite{visual6502source}. Each transistor has three terminals (gate,
source, drain), but the methods of the original dataset were unable to
consistently resolve which of the C1 and C2 terminals were source and
drain, leading to ambiguity in our encoding. We identified a region
consisting of three registers X, Y, and S via visual
inspection and focused our efforts there. We created a total of six
connectivity matrices by examining possible terminal pairings. One
graph, for example, $R^{gc_1}(e_i, e_j)=1$ if transistor $e_j$ and
$e_j$ are connected via pins $g$ and $c_1$.

\newpage
\section*{Supplemental Material}

\subsection*{Other Likelihoods}
\label{supp:otherlikelihoods}

We reparameterized the Logistic-Distance Bernoulli likelihood to
better capture the microprocessor data structure. We are explicitly
setting the maximum probability $p$ of the logistic function on a
per-component basis, drawing from a global $p \sim \operatorname{Beta}(\alpha_{hp},
\beta_{hp})$. Then $\lambda$ is set for each component as a global
hyperparameter, $\lambda$.

The ``logstic-distance Poisson'' spatial model is used to explicitly
mode the count of synapses, $c$, between two neurons. The probability
of c synapses between two neurons is distributed $c \sim
\textrm{Poisson}(c | r)$, where $r$ (the ``rate'') is generated by a
scaled logistic function (the logistic function has range $[0,
1]$. For each component $\eta_{mn}$ we learn both the threshold
$\mu_{mn}$ and the rate scaling factor $r_{mn}$ Thus if for cells $m$
and $n$ are likely to have on average $20$ synases if they are closer
than $5 \mu m$, then $\mu_{mn} = 5$ and $r_{mn} = 20$ "

Thus the probability of $R(e_i, e_j) = c$ synapses between two cells $e_i$ and $e_j$ is given by:
\begin{eqnarray}
r^* &=& \frac{1.0}{1 + \exp \frac{d(e_i, e_j) - \mu_{mn}}{\lambda}}\\
r & = & r^* \cdot (r_{mn} - r_{min}) + r_{min} \\
R(e_i, e_j) \sim \textrm{Poisson}(c | r)
\end{eqnarray}

where $\lambda$ and $r_{min}$ are per-graph parameters. Per-component parameters $\mu_{mn} \sim \exp(\mu | \mu^{hp})$ and $r_{mn} \sim \exp(r_{mn} | r_{scale}^{hp})$. 

\subsection*{Source code and data}

All source code and materials for running experiments can be
obtained from the project website, at \\

\href{http://https://github.com/ericmjonas/netmotifs}{https://github.com/ericmjonas/netmotifs/}

and the content of this paper along with scripts to run
experiments and generate all figures can be found at

\href{http://https://github.com/ericmjonas/netmotifs}{https://github.com/ericmjonas/connect-disco-paper/}

All preprocessed data has been made publicaly available as well. 

Please contact the author for pre-publication access. 

\subsection*{Extension to multiple graphs}
\label{supp:multigraph}
The model can handle multiple graphs $R^q$ simultaneously with a shared clustering by extending the likelihood to include the product of the likelihoods of the individual graphs. 

\begin{equation}
  p(\vec{c}, \{\eta^q\}, \{\theta^q\} | \{R^q\} ) \propto \prod_q \Bigg(\prod_{i, j} p(R^q(e_i, e_j) | f(d(e_i, e_j) | \eta^q_{c_ic_j}, \theta^q) \prod_{m, n} p(\eta^q_{mn} | \theta^q)  p(\theta^q) \Bigg) p(\vec{c} | \alpha) p(\alpha) 
\end{equation}

\FloatBarrier
\subsection*{Hyperprior grids and hyperprior inference}
\label{supp:hyperpriors}

For the mouse retina Logistic-Distance Bernoulli model, we gridded
$\mu^{hp}$ and $\lambda^{hp}$ into 40 $\log_{10}$-spaced points 1.0
and 80. 

For the c. elegans data with the Logistic Distance poisson model, we
gridded $\mu_{hp}$ and $\lambda$ into 20 $\log_{10}$-spaced points
bween 0.2 and 2.0, and the $ratescale^{hp}$ parameter into 20
$\log_{10}$-spaced points between 2.0 and 20.0. We globally set
$rate_{min}=0.01$.

For the microprocessor with the Logistic Distance fixed lambda
Bernoulli likelihood, we gridded $mu_{hp}$ into 50 $\log_{10}$-spaced
points between 10 and 500 and set $\lambda=\mu_{hp}/10$. $p_{min} \in
\{0.001, 0.01, 0.02\}$ and both $p_\alpha$ and $p_\beta \in \{0.1,
1.0, 2.0\}$.

\section*{Measuring clustering similarity}

The adjusted rand index (ARI) is a measure of the similarity of two
different clusterings \autocite{Hubert1985} -- two identical clusters
have an ARI of 1.0 while progressively more dissimilar clusters have
lower ARIs, becoming negative as the clustering gets anti-correlated.

\end{document}